\documentclass[onecolumn,11pt]{article}
\usepackage[margin=2.2cm,a4paper]{geometry}
\pdfoutput=1
\usepackage[utf8]{inputenc}

\usepackage{amsmath,amsthm}
\usepackage{bm}  
\usepackage{verbatim}
\usepackage{xcolor}

\usepackage[backend=biber,sorting=none,style=phys-jmr,biblabel=brackets,backref=false,bibencoding=utf8,maxnames=20,eprint=true]{biblatex}
\makeatletter
\pretocmd{\blx@head@bibintoc}{\phantomsection}{}{\ddt}
\makeatother
\DefineBibliographyStrings{english}{%
  backrefpage = {page},
  backrefpages = {pages},
}
\usepackage{titlesec}

\titleformat*{\section}{\bfseries}
\titleformat{\subsection}[runin]{\normalsize\bfseries}{\thesubsection. }{0.0em}{}[]
\titleformat*{\subsubsection}{\bfseries}
\titleformat*{\paragraph}{\large\bfseries}
\titleformat*{\subparagraph}{\large\bfseries}

\titlespacing\section{0pt}{12pt plus 4pt minus 2pt}{2pt plus 2pt minus 2pt}

\usepackage[bookmarks,colorlinks,breaklinks]{hyperref}  
\definecolor{dullmagenta}{rgb}{0.4,0,0.4}   
\definecolor{darkblue}{rgb}{0,0,0.4}
\hypersetup{linkcolor=red,citecolor=blue,filecolor=dullmagenta,urlcolor=darkblue} 

\usepackage[bitstream-charter,greekuppercase=italicized,cal=cmcal]{mathdesign}
\usepackage[T1]{fontenc}

\usepackage{amsbsy,verbatim,graphicx}
\usepackage{units}

\newcommand{\ket}[1]{|#1\rangle}

\newcommand{\bra}[1]{\langle #1|}

\newcommand{\ketbra}[1]{\ket{#1}\bra{#1}}

\newcommand{\tr}{{\text{Tr}}}

\def\up{{\uparrow}}
\def\down{{\downarrow}}

\DeclareFontFamily{U}{bbold}{}
\DeclareFontShape{U}{bbold}{m}{n}
 {
  <-5.5> s*[1.069] bbold5
  <5.5-6.5> s*[1.069] bbold6
  <6.5-7.5> s*[1.069] bbold7
  <7.5-8.5> s*[1.069] bbold8
  <8.5-9.5> s*[1.069] bbold9
  <9.5-11> s*[1.069] bbold12 
  <11-15> s*[1.069] bbold12
  <15-> s*[1.069] bbold17
 }{}

\DeclareRobustCommand{\id}{%
  \text{\usefont{U}{bbold}{m}{n}1}%
}

\newtheorem*{parse}{Parsing rule}
\newtheorem*{context}{Context rule}

\usepackage{setspace}
\usepackage{titling}

\posttitle{\par\end{center}}
\predate{}
\postdate{}
\begin{document}
\author{{\normalsize Joseph M.~Renes}\\
{ \normalsize \emph{Institute for Theoretical Physics, ETH Z\"urich, 8093 Z\"urich, Switzerland}}
}

\title{\large {\bf Consistency in the description of quantum measurement: \\ Quantum theory can consistently describe the use of itself}}

\date{\vspace{-\baselineskip}}


\maketitle

\begin{abstract}
Recent extended formulations of the Wigner's friend thought experiment throw the measurement problem of quantum mechanics into sharper relief. 
Here I respond to an invitation by Renner to provide a consistent and concrete set of rules for quantum mechanics which can avoid the apparent paradox formulated by Frauchiger and Renner [Nat.\ Comm.\ {\bf 9}, 3711 (2018)]. 
I propose a slight addition to standard textbook quantum mechanics, in the form of two rules, which avoids the paradox.
The first specifies when a given quantum dynamics can be interpreted as a measurement. 
Potentially any dynamics can, but doing so depends on the context of other performed operations. 
The second requires that a joint context be used to determine whether several different dynamical evolutions can all be interpreted as measurement. 
The paradox is then avoided because not every participant in the Frauchiger-Renner thought experiment regards the stated dynamical descriptions of the experiment as measurements. 
\end{abstract}
\vspace{0.5\baselineskip}

\section{Introduction}

The measurement problem in quantum mechanics is the apparent inconsistency between two possible descriptions of measurement. 
On the one hand, according to the projection postulate, measurement of a system $S$ is described by a set of projection operators $P_S(k)$ such that the result of the measurement on a system in quantum state $\ket{\psi}_S$ is the value $k$ with probability $p(k)=|\,{}_S\!\bra\psi P_S(k)\ket{\psi}_S|^2$ and the state collapses to $P_S(k)\ket{\psi}_S/\sqrt{p(k)}$. 
On the other hand, measurement is a dynamical process, and so surely has a dynamical description in terms of the Schrödinger equation for a suitable interaction Hamiltonian between the system and the measurement apparatus. 
The former description is called Process 1 (P1) by von Neumann and the latter Process 2 (P2)~\cite{von_neumann_mathematische_1932}. 

The distinction is exemplified in the Wigner's friend thought experiment~\cite{wigner_remarks_1963,wigner_problem_1963}. According to the projection postulate, if Wigner's friend makes a measurement of a spin-$\nicefrac 12$ system prepared in the $+\hat x$ eigenstate in the $\hat z$ basis, i.e. the state $\ket \up_S+\ket \down_S$ in the basis $\{\ket{\up},\ket{\down}\}$, the result $\up$ or $\down$ will be random and the state of the spin will be projected to the corresponding state $\ket\up_S$ or $\ket \down_S$. 
But Wigner, observing the process from outside, can use the Schrödinger equation to describe the measurement process, which results in the joint state of the spin and the friend $F$ in the entangled superposition state $\ket{\up}_S\otimes \ket{\text{``up''}}_{F}+\ket{\down}_S\otimes \ket{\text{``down''}}_{F}$. 
Here $\ket{\text{``up''}}_F$ denotes the state of the friend having observed spin up, and similarly for $\ket{\text{``down''}}_F$.  
Which description is correct?

Recently a more elaborate version of the thought experiment which gives rise to a logical paradox has been put forth by Frauchiger and Renner (FR)~\cite{frauchiger_quantum_2018}. 
Their paradox arises when several observers each use quantum mechanics to describe the rest of the world, inluding the other observers. Making inferences about what other observers must have observed given the precise setup of the experiment (and the rules of quantum mechanics), the observers come to a conclusion about what the result of the final measurement in the experiment must be. 
However, in at least some rounds of the experiment, the prediction will contradict the actual outcome. 
In subsequent discussion of the result, Renner challenged those disputing the result or the implications of the result to come up with consistent rules of quantum mechanics such that computers programmed with the rules would not run into contradictions~\cite{renner_comment_2018}. 
He points out that it is insufficient to simply give general reasons why their formulation breaks down, because to do so is effectively to admit that the rules of quantum mechanics are not precise enough to properly address the FR scenario. 
Hence some kind of patch is required. 

Here, I propose a rule for parsing descriptions of quantum dynamics as measurement and a rule for taking over inferences made by other observers which meets this challenge. 
The proposed rules are to be understood as a very slight addition to the usual rules of quantum theory set out by Dirac~\cite{dirac_principles_1967} and von Neumann~\cite{von_neumann_mathematische_1932}. 
Under the proposed parsing rule, \emph{every} dynamical evolution can potentially be interpreted as a measurement. 
But the extent to which this is possible is determined by the context of other operations, including other measurements, which take place. 
This will resolve the paradox in that several of the computers in the thought experiment will not recognize various parts of the dynamical description of the experiment as measurements, and therefore the logical chain of reasoning which leads to the paradox will be broken. 
In the next section I formulate the two rules precisely and illustrate them with several simple examples including the quantum eraser. 
Then I discuss the resolution of the FR paradox in detail, and then turn to the broader implications of the proposed rules, as well as the relationship to other approaches.  

\section{Parsing quantum dynamics as measurement}
\label{sec:rules}
In a dynamical description of measurement, the interaction Hamiltonian induces a unitary transformation $U_{SM}$ on the system $S$ and the measurement device $M$. 
As the initial state of the measurement device is taken to be in a fixed state, call it $\ket 0_M$, we can just as well represent the dynamical description by an isometry $V_{SM|S}=U_{SM}\ket{0}_M$, which maps the state space of $S$ to that of $S$ and $M$. 
For simplicity, denote by $R$ the joint system $S$ and $M$. 
Indeed, an isometry $V_{R|S}$ is the most generic description of any kind of quantum dynamics, as ensured by the Stinespring representation~\cite[\S8.2.2]{nielsen_quantum_2010}\cite[\S2.2]{watrous_theory_2018}. 
Now we can regard $R$ as a generic system, possibly but not necessarily containing $S$. 
This description also captures the possibility that the dynamics transforms $S$ into some other output system $S'$, as we can regard $R$ as containing $S'$. 
Meanwhile, the most general statistical description of a measurement is as a POVM~\cite[\S2.2.6]{nielsen_quantum_2010}\cite[\S2.3]{watrous_theory_2018}, a collection of positive operators $\Lambda_S(k)$ on the system $S$ which sum to the identity operator: $\sum_k \Lambda_S(k)=\id_S$. 

Now consider some further set $\mathcal C$ of quantum operations of any kind, be they actions on $S$, $R$, or anything else.
Then I propose the following rule for deciding when a given dynamics as specified by an isometry can be understood as a measurement. 
\begin{parse}
An isometry $V_{R|S}$ can be understood as a dynamical description of a POVM $\{\Lambda_S(k)\}$ within the context of $\mathcal C$ if there exists an observable $A_R$ with spectral projections $P_R(k)$ such that 
\begin{enumerate} 
	\item $\tr_R[P_R(k) \,V_{R|S} \rho_{SE} V_{R|S}^*] = \tr_S[\Lambda_S(k)\, \rho_{SE}]$ for all states $\rho_{SE}$ involving $S$ and any other system $E$, and 
	\item All operations in $\mathcal C$ which occur after the action of $V_{R|S}$ commute with $A_R$. 
\end{enumerate}
\end{parse}
The observable $A_R$ is, essentially, the measurement result, the degrees of freedom in which the measurement result is stored. 
The first condition ensures consistency between the P1 and P2 descriptions of measurement. 
Note that this condition must hold for all quantum states, possibly entangled with any other system $E$, not just the particular state appearing in any given experimental setup to be analyzed. 
The second condition requires that the measurement result not be disturbed by further operations in the context $\mathcal C$, but only in this context.  
A particular dynamics may parse as measurement in one context and not in another.
That is, \emph{whether or not a particular interaction constitutes a measurement is not absolute in this framework.}

The parsing rule refers to POVMs and not quantum instruments, which in addition to the statistics of the measurement outcome, also specify the post-measurement state. In the case of rank-one projective measurements, the POVM implies the instrument description, but this is not true for general POVMs. 
For a general POVM with elements $\Lambda_S(k)$, a possible instrument description is given by a decomposition of the $\Lambda_S(k)$ into measurement operators $M_{S'|S}(k,\ell)$ with $\sum_\ell M_{S'|S}(k,\ell) M_{S'|S}(k,\ell)^*=\Lambda_S(k)$; the post-measurement state for outcome $k$ is then, up to normalization, just $\sum_\ell M_{S'|S}(k,\ell)\rho_S M_{S'|S}(k,\ell)^*$ for input state $\rho_S$. 
Observe that the instrument description is handled by the flexibility in the isometry: simply pick $V_{S'ML|S}=\sum_{k,m} M_{S'|S}(k,\ell)\otimes \ket{k}_{M}\otimes \ket{\ell}_{L}$. 

To handle several measurements, I propose a separate rule. 
Essentially, contexts have to be combined when parsing several isometries as measurements.
\begin{context}
Suppose that isometries $V^{(1)},\dots, V^{(n)}$ can each individually be regarded as measurements $M_1,\dots, M_n$ according to the parsing rule with contexts $\mathcal C_1,\dots, \mathcal C_n$, respectively. 
Then the entire collection of isometries can be regarded as the describing the collection of measurements only if $V^{(k)}$ parses as measurement $M_k$ in the joint context $\mathcal C_1\cup\dots\cup \mathcal C_n$. 
\end{context}
\noindent According to this rule, it is not enough for, say, all pairs of isometries to be interpretable as measurements in order to interpret the entire collection of isometries as describing the entire collection of measurements. 
Requiring the use of the joint context ensures that the measurement result operators from all the measurements form a commuting algebra of observables. 
Therefore propositions about their simultaneous values form a Boolean algebra (see e.g.~\cite[Lemma 1.1]{wilce_quantum_2017}). 
This is the setting necessary to carry out inferences of some measurement results given other measurement results.

It is worth explicitly recalling the motivation at this point, of formulating rules suitable for use by computers in performing the FR thought experiment. 
This setting invites us to regard quantum theory as any other tool for predicting the outcomes of experiments, particularly as just a special kind of probabilistic model, which are widely used in statistics, machine learning, and computer science. I will return to the interpretational issues in \S\ref{sec:Discussion}. 

\section{Wigner's friend and quantum eraser}

Let us now return to the original Wigner's friend thought experiment from the introduction, but phrase it more directly in usual qubit language. 
The measurement in question is a projective measurement in the basis $\{\ket 0_S ,\ket 1_S\}$, for which a possible dynamical description is given by the \textsc{cnot} gate applied from $S$ to an ancilla qubit $M$ initially in the state $\ket{0}_M$. 
That is, system $S$ controls the not gate applied to $M$.  
For an initial state $\ket{\psi}_S=\alpha\ket{0}_S+\beta\ket{1}_S$, the output of the \textsc{cnot} is simply $\ket{\psi'}_{SM}=\alpha\ket{00}_{SM}+\beta\ket{11}_{SM}$. 
According to the proposed rule, this is a perfectly valid description of a $(\sigma_z)_S$ measurement, provided that all the subsequent interactions Wigner intends to take on the friend and the qubit commute with $A_R=\id_S\otimes (\sigma_z)_M$. 
For instance, Wigner may ask what the result was. 
Call this context $\mathcal C_{Z}$.
This operation can be modeled as a projective measurement of $M$, or by a further outside observer as another \textsc{cnot} gate, this time with the control on $M$ and the target Wigner's memory. 

As an example that the \textsc{cnot} gate either can or cannot be interpreted as a measurement, depending on the context, consider two different further contexts of operations: $\mathcal C_{XZ}$, which builds on $\mathcal C_Z$ by Wigner additionally measuring $S$ in the $\ket{\pm}_S$ basis, and $\mathcal C_{XX}$, in which he subsequently measures both $S$ and $M$, each in the basis $\ket{\pm}$.
According to the proposed rule, the \textsc{cnot} description of measurement is valid for $\mathcal C_{XZ}$, but not for $\mathcal C_{XX}$. 
The former context does not add anything which does not commute with the choice of $A_R$ for $\mathcal C_Z$, while for the latter there is simply no operator $A_R$ which satisfies both conditions of the rule. 
Wigner's actions are all in the $\sigma_x$ basis, while $A_R$ must deal with the $\sigma_z$ basis. 

In fact, this example is precisely the quantum eraser~\cite{scully_quantum_1982,scully_quantum_1991,englert_remarks_1999,englert_quantum_1999}, a wonderful irony of quantum mechanics in which a putative measurement (dynamically described by the \textsc{cnot} gate) is in some sense undone by a subsequent measurement of the ancilla qubit. 
If we regard the $(\sigma_x)_S$ observable as indicating the phase relationship of two paths of a single photon in a Mach-Zehnder interferometer, and $(\sigma_z)_S$ as indicating the path information, then the $\mathcal C_{XZ}$ context is compatible with the \textsc{cnot} gate as performing a $\sigma_z$ measurement. 
The phase relationship is completely randomized, just as when we use the projective description of the path measurement.  
However, a subsequent (phase) measurement of $M$ restores the phase relationship in $S$.
This can be seen by examining the state after the application of the \textsc{cnot} gate more closely:
\begin{align}
\ket{\psi'}_{SM}
&=\tfrac{1}{\sqrt{2}}\alpha\ket{0}_S(\ket{+}+\ket{-})_M+\tfrac{1}{\sqrt{2}}\beta\ket{1}_S(\ket{+}-\ket{-})_M\\
&=\tfrac{1}{\sqrt{2}}\ket{\psi}_S\ket{+}_M+\tfrac{1}{\sqrt{2}}(\sigma_z)_S\ket{\psi}_S\ket{-}_M\,.
\end{align}
Applying a controlled-phase operation from the $\ket{\pm}_M$ basis to $S$ will remove the $\sigma_z$ in the $\ket{-}_M$ term, restoring the initial state. 
Hence, in the context $\mathcal C_{XX}$, the \textsc{cnot} gate cannot be regarded as a measurement of $(\sigma_z)_S$, for in this description the coherence of the input state remains.

One more related context nicely illustrates some of the subtleties of the parsing rule. 
Suppose instead that Wigner subsequently measures $M$ in the basis $\ket{\pm}_M$ and $S$ in the basis $\ket{k}_S$; call this $\mathcal C_{ZX}$. 
Now we can no longer use the above choice of $A_R$, with the measurement result stored in the memory $M$. 
However, we can take the measurement result observable to be $A_R=(\sigma_z)_S\otimes \id_M$. 
This is valid for both $\mathcal C_{ZX}$ and $\mathcal C_{Z}$ (though not $\mathcal C_{XZ})$. 
Here the \textsc{cnot} gate is meant to describe measurement of $(\sigma_z)_S$, and subsequently Wigner measures $(\sigma_z)_S$ himself, along with $(\sigma_x)_M$. 
We regard the measurement result as stored in $S$.
Indeed, for an input state $\ket{\psi}_{S}=\alpha\ket 0_S+\beta\ket1_S$, when Wigner checks the value of $(\sigma_z)_S$, the measurement probabailities are precisely as they would be in a projective description of the measurement, namely $|\alpha|^2$ for 0 and $|\beta|^2$ for 1. 
This also holds if the input state is maximally entangled with an additional qubit $E$;  Wigner's $(\sigma_z)_S$ measurement will be perfectly corrleated with $(\sigma_z)_E$.  
For its part, the $(\sigma_x)_M$ results are simply random for either input state. 
Note that this fact is not compatible with a projective description of the measurement of $(\sigma_z)_S$, but the proposed rule does not require agreement of further measurements on $M$. Nor should it have to: Not every physical implementation of a given measurement is expected to have the same effect on the measurement apparatus, the surrounding environment, and so forth.




\section{A resolution to the FR paradox}


In the ``extended Wigner's friend'' thought experiment of FR there are two friends ostensibly making measurements and two outside observers. 
Here we follow the naming convention and notation of~\cite{nurgalieva_testing_2020}, where the friends are named Alice and Bob, and the corresponding outside observers Ursula and Wigner. 
In the thought experiment, Alice is meant to measure a qubit system $R$ in the computational basis, Bob a qubit system $S$, also in the computational basis. 
These measurements are described dynamically by the isometries 
\begin{align}
V_{RA\bar A}:\ket{k}_R\mapsto \ket{\text{lab}_k}_{RA\bar A}:=\ket{k}_R\ket{\text{``I observed outcome $k$''}}_A \ket{\text{env}_k}_{\bar A}
\end{align}
for Alice, and similiarly $V'_{SB\bar B}$ for Bob. 
Here the states $\ket{\text{``I observed outcome $k$''}}_A$ are orthogonal states of Alice for $k=0,1$, while the states $\ket{\text{env}_k}_{\bar A}$ are arbitrary states of the environment in Alice's laboratory and need not be orthogonal. 
Subsequent to these operations, Ursula and Wigner measure the state of Alice's and Bob's laboratories, respectively. 
In particular, Ursula measures in the basis 
\begin{align}
\ket{\text{lab}_\pm}_{RA\bar A}:=\tfrac1{\sqrt2}(\ket{\text{lab}_0}_{RA\bar A}\pm \ket{\text{lab}_1}_{RA\bar A}\,,
\end{align}
and Wigner measures Bob's laboratory similarly.

Let us take the actions taken in the thought experiment to be formally specified by $V_{RA\bar A|R}$, then $V'_{SB\bar B|S}$, and subsequent projective measurements by Ursula and Wigner. 
Furthermore, the initial state of $R$ is given by $\sqrt{\nicefrac 13}\ket{0}_R+\sqrt{\nicefrac23}\ket 1_R$, while the initial state of $S$ is generated by Alice using the isometry $U_{SA|A}$ which has the action 
\begin{align}
\ket{\text{``I observed outcome $0$''}}_A &\mapsto \ket{\text{``I observed outcome $0$''}}_A\ket{0}_S\,,\\
\ket{\text{``I observed outcome $1$''}}_A &\mapsto \ket{\text{``I observed outcome $1$''}}_A\ket{+}_S\,.
\end{align}
This fully formally specifies the actions taken in the thought experiment. 
The issue is whether the isometric actions $V_{RA\bar A|R}$ and $V_{SB\bar B|S}$ constitute measurements. According to the proposed rule, there is no absolute answer to this question, but nevertheless it will allow us to avoid the FR paradox. 

The paradox arises from what the various parties infer are the measurement results of certain other parties. 
After the first step in the thought experiment, Alice may infer what she believes will be Wigner's measurement result, given her measurement result. 
To make sense of such a statement, a computer playing the role of Alice will have to parse $V_{RA\bar A|R}$ as a description of her measurement in the context of her preparation of $S$, Bob's measurement, and finally Wigner's subsequent measurement of $SB\bar B$. 
Otherwise she will state that the description of the experiment does not accord with the actions she is taking. 
This particular situation is unproblematic, though: All of the operations commute with the observable corresponding to her measurement result, which is diagonal in the ``I observed...'' basis. 
Hence, if Alice observes outcome 1, then she is certain the state delivered to Bob will be $\ket{+}_S$ and therefore Wigner's measurement result will be $+$. 

After Bob's measurement in the next step in the thought experiment, he attempts to indirectly infer Wigner's measurement result, by inferring Alice's measurement result and then making use of her inference of Wigner's result.  
The Bob computer will parse the isometric description of his operation as a valid description of his measurement in the context of Alice's isometric actions, as these occur prior to his measurement. 
And so he will infer that if the measurement result he obtained was 1, then Alice's measurement result was likewise 1. 
Similarly, the Bob computer will parse the description of Alice's and Wigner's operations as measurements and come to the same conclusion about Wigner's outcome conditioned on Alice's result. 
As Alice does not directly tell him her inference, it is necessary for him to repeat this calculation.  
However, according to the parsing rule, the isometry $V'_{SB\bar B|S}$ cannot be regarded as describing Bob's measurement in the context of Wigner's subsequent measurement (or any set of actions which contains it). 
He can make sense of $V$ as describing Alice's measurement both in the context of his own actions as described by $V'$ and in the context of Wigner's measurement, but $V'$ does not parse as a measurement when including Wigner's measurement in the context. 
By the context rule he is therefore precluded from regarding the entire formal description as a  description of measurements by Alice, himself, and Wigner. 
Hence he lacks a Boolean algebra of propositions about the outcomes of all three measurements in which to carry out the inference of Wigner's result given his result. 
So this route to inferring Wigner's measurement result is not available. 
At this point the chain of reasoning toward the paradox is broken.

Note that the chain of reasoning would be broken by Alice if she were programmed to include Ursula's measurement in the context when predicting Wigner's outcome. 
Doing so is not necessary to make the logical inference, and so we did not include it. 
However, there is no avoiding Bob's need to include Wigner. 
He has to construct the setting in which he, Alice, and Wigner measure to make any sense of using  her inference of Wigner's result, in accordance with the context rule. 
Hence this rule is sufficient to avoid the paradox. 
Since the thought experiment only involves combining the Alice-Bob and Alice-Wigner contexts, the  paradox could be viewed as evidence that the context rule is also necessary, but I do not pursue this further here. 




\section{Discussion}
\label{sec:Discussion}

At a broad conceptual level, the proposal here is not particularly novel. 
Many have made suggestions for how the FR paradox can be avoided which are similar in spirit. 
Closely related are
the discussion of ``stable versus relative facts'' by Di Biagio and Rovelli~\cite{di_biagio_stable_2021}, the principle of ``superpositional solipsism'' of Narasimhachar~\cite{narasimhachar_agents_2020}, and the criteron of Elouard \emph{et al.}\ that observers are only observers when they can store the measurement result in some degree of freedom until the end of the experiment~\cite{elouard_quantum_2021}. 
Elouard \emph{et al.}\ also point out the relevance of the quantum eraser to the discussion of FR paradox, as do Żukowski and Markiewicz~\cite{zukowski_physics_2021}.
The proposal here also has a lot in common with the consistent histories approach to quantum mechanics~\cite{griffiths_consistent_1984,gell-mann_quantum_1997,omnes_logical_1988,griffiths_consistent_2002,griffiths_consistent_2014}, about which more will be said below. 
Here the main contribution, and distinction to previous work, is to give a concise and concrete set of additional rules to the standard axioms of Dirac and von Neumann which formally instantiate many of the conceptual objections to the FR paradox.
My aim is to be as conservative as possible, to augment the rules just enough to resolve the paradox by stating that the description does not accord with the notion of measurement in certain contexts, but not so much as to invalidate the unitary description of the original Wigner's friend scenario.

The FR paradox is formulated as a theorem that three reasonable-sounding assumptions Q, S, and C lead to a contradiction~\cite[Theorem 1]{frauchiger_quantum_2018}. 
Since the parsing and context rules avoid the paradox, it is interesting to look at which assumption is violated.
Roughly speaking, Q stands for quantum theory, i.e.\ predictions are to be made using quantum theory, S for single outcomes, that the parties each observe single outcomes in their measurements, and C consistency, that their predictions be consistent and that one party can take over the inferences of another. 
Although the proposed rules are a slight modification of Q, the bigger violation is to C.  
Since measurements are not regarded as absolute, it is not always possible to take over inferences of other parties. 
This is certainly an uncomfortable state of affairs, as in principle the fact of the matter as to whether a given experiment had a given result is now relative to other operations, including operations in the future. 
This conclusion is perhaps not so dramatic in the usual setting of the quantum eraser, where the ancilla target of the \textsc{cnot} gate is itself a qubit. 
Replacing the qubit with an observer friend is the uncomfortable step, though nothing in our current formulation of quantum theory forbids it. 


That dropping absoluteness of measurements from FR's Assumption C avoids the paradox has already been pointed out by several authors~\cite{healey_quantum_2018,debrota_respecting_2020,di_biagio_stable_2021,elouard_quantum_2021,cavalcanti_view_2021}. 
Several recent no-go theorems, in the spirit of the Bell inequalities, also call absoluteness of measurement into question~\cite{brukner_quantum_2017,brukner_no-go_2018,bong_strong_2020,cavalcanti_implications_2021}.
However, dropping absolutness raises another problem, for then it is unclear if one party can reason about another party at all. 
The contribution here is to describe very precisely when the parties in the thought experiment can and cannot regard the actions of the others as measurements, and hence reason about their results.




It has been my intention to avoid discussion of interpretations of quantum mechanics in formulating the framework of the parsing and context rules.
However, this is to some extent impossible, and it is worthwhile to briefly examine the connections between the approach here and various interpretations.
As alluded to at the end of \S\ref{sec:rules}, the approach here fits naturally into the view of quantum theory as a generalized probability theory, ultimately as means to determine probabilities of outcomes of experiments. 
I conceive of it most directly in terms of what Werner refers to as the ``minimal statistical interpretation''~\cite{werner_uncertainty_2019}. 
This is belied by the language used, e.g. ``The Bob computer will parse the isometric \emph{description} of his operation...''.
It is also related to the Copenhagen and ``neo-Copenhagen''~\cite{de_muynck_towards_2004,brukner_quantum_2017} interpretations, QBism~\cite{caves_subjective_2007,debrota_respecting_2020}, as well as Bub's information-theoretic approach~\cite{bub_defense_2020}.  
Separately, taking the view that measurements are not absolute accords with relational quantum mechanics~\cite{rovelli_relational_1996,di_biagio_stable_2021}, along with QBism and neo-Copenhagen. 
At the structural level, the context rule is assuredly related to the requirement of only considering consistent histories in that approach, as is the focus on having a Boolean algebra of propositions. 
The focus on algebras of measurement events as commutative (and therefore Boolean) is also stressed in the ETH approach~\cite{frohlich_quantum_2015,frohlich_brief_2020} and by Brukner~\cite{brukner_no-go_2018}.
However, both consistent histories and the ETH approach appear to differ completely from the present approach in the two aforementioned points. 

Another way to understand the similarities and differences is that the present approach retains the P1 description of measurement. 
To do so requires the Heisenberg cut~\cite{heisenberg_wandlungen_1934,heisenberg_ist_1935,bacciagaluppi_heisenberg_2009}, i.e.\ a boundary between observer and object across which P1 is used to describe measurement. 
However, the cut is subjective: One observer's P1 description is another (outside) observer's P2 description.     
The parsing rule uses consistency between P1 and P2 as the criteria for accepting a given dynamics as a description of a measurement.
I take it as a crucially important fact about quantum mechanics that these two descriptions can be consistent, and that ultimately favoring one over the other in an interpretation would be to lose something vital. 
In any case, the parsing and context rules do not resolve the measurement problem, they merely shift it to the somewhat uncomfortable notion that measurements are not absolute. 
I do not claim that this is necessarily an aesthetically or philosophically pleasing way to understand quantum mechanics, merely that it is consistent and avoids the FR paradox.



\subsection*{Acknowledgements} Many thanks to Christopher A.\ Fuchs, Christophe Piveteau, Renato Renner, L\'idia del Rio, and Henrik Wilming for useful discussions. I am especially grateful to V.\ Vilasini and Nuriya Nurgalieva for careful reading of the manuscript. I acknowledge financial support from the Swiss National Science Foundation via the National Center for Competence in Research for Quantum Science and Technology (QSIT). 

\printbibliography[heading=bibintoc,title={\normalsize References}]

\end{document}